\documentclass[aps,pra,twocolumn,showpacs,groupedaddress,amsmath,amssymb]{revtex4} 

\usepackage{graphicx}
\usepackage{dcolumn}
\usepackage{bm}
\usepackage{hyperref}
\usepackage{color}

\begin{document}
	
	\title{Additional complex conjugate feedback induced explosive death and multistabilities } 
	\author{K. Sathiyadevi$^1$, D. Premraj$^2$, Tanmoy Banerjee$^3$, and M. Lakshmanan$^4$}
	\address{$^1$Complex Systems and Applications Lab, Rajalakshmi Institute of Technology, Chennai, 600124, Tamilnadu, India.  \\
		$^2$Centre for Nonlinear Dynamics, Chennai Institute of Technology, Chennai-600 069, Tamilnadu, India.  \\ $^3$Chaos and Complex Systems Research Laboratory, Department of Physics, University of Burdwan,  Burdwan 713 104, West Bengal, India.\\
		$^4$Department for Nonlinear Dynamics, School of Physics,   Bharathidasan University, Tiruchirappalli 620024, Tamilnadu, India 
	}
	\received{:to be included by reviewer}
	\date{\today}
	\begin{abstract}
		Many natural and man-made systems require suitable feedback to function properly. In this study, we aim to investigate the impact of additional complex conjugate feedback on globally coupled Stuart-Landau oscillators. We find that this additional feedback results in the onset of symmetry breaking clusters and out-of-phase clusters. Interestingly, we also find the existence of explosive amplitude death along with disparate multistable states. We characterize the first-order transition to explosive death through the amplitude order parameter and show that the transition from oscillatory to death state indeed shows a hysteresis nature. Further, we map the global dynamical transitions in the parametric spaces. In addition, to understand the existence of multistabilities and their transitions, we analyze the bifurcation scenarios of the reduced model and also explore their basin stability. Our study will shed light on the emergent dynamics in the presence of additional feedback.		
	\end{abstract}
	
	\maketitle 
	
	\section{Introduction}
	\label{sec:intro}
	\par In many natural and man-made systems including neural networks, vision systems, lasers, etc., feedback is essential for proper functioning and it can be used for enhancing the performance of such systems \cite{r1_n,r2_n,vis,laser}. { Besides,  several investigations have been conducted using various nonlinear models to get a better understanding on the dynamical characteristics as well as  the impact of feedback \cite{feed1,rev1,rev2,rev3,rev4,rev5}.}  Recent investigations reveal that appropriate feedback can restore oscillatory dynamics or induce the oscillation death state depending on the properties of the ensembles, coupling architecture, and the kind of feedback and its strength \cite{2r1,2r2,2r3}.  Importantly, among the various intrinsic and extrinsic parameters such as time-delay, low-pass filtering, mean-field density, etc., typically, the feedback is used for controlling purposes \cite{prem1,prem2,prem3,tan_mean1,tan_mean2,tan_mean3,tan_filt1,pon_filt,sat_uday1}.  For instance,  the conjugate feedback or the self-feedback approach is used for controlling the  birhythmicity in different realistic models such as energy harvesting systems as well as biochemical systems \cite{tan_bi1,tan_bi2}. The self feedback factor can also be used for controlling the spontaneous symmetry breaking oscillations \cite{sat_pon}. The linear feedback in the diffusion term can cause suppression of aging whereas   by providing the feedback in the mean-field term one can enhance  the aging region \cite{sat_ag1,sat_ag2,sa}. The robustness of dynamical activity has also been demonstrated in damaged scale-free and small-world networks using external feedback mechanisms \cite{diba_res}.  
	\par On the other hand, explosive transitions like explosive percolation (EP), explosive synchronization (ES), and  explosive death (ED) have also been extensively studied in networks of coupled systems by various research groups \cite{sarika1,sarika2,exp_perc1,perc2}.  Each of these transitions has its own merits and demerits. Among all of them, explosive death is an interesting phenomenon that exhibits a first-order jump from oscillatory state to oscillation death state \cite{expl}.  Understanding the reason behind the emergence of such ED may help to predict or prevent them earlier in real-time instances.  Originally, the explosive death was reported in frequency weighted coupling under three different frequency distributions \cite{expl}. Followed by ED also investigated mean-field diffusive coupling and it  is reported that the occurrence of ED in van der Pol and Lorenz oscillators \cite{ed2mean}. Later, it was also demonstrated in the coupled limit cycle and chaotic oscillators under different kinds of network topologies including nearest neighbors, all to all, nonlocal, and star network connectivities \cite{ed3env,ed4star}. { Very recently, conjugately coupled van der Pol oscillators induced ED (first order transitions during the forward and backward transitions) and semi-ED (first order transitions occur during forward transitions, while second-order transitions occur during backward transitions, or vice versa.)  have also been demonstrated \cite{semi_ex}.} Further, the dynamical mean-field interaction induced ED reported under limit cycle, chaotic, and neural network \cite{dyn}. As of now, the explosive death state has been identified due to various kinds of interactions and frequency distribution.  In this paper, \textit{we examine whether an additional complex conjugate feedback can exhibit explosive transitions and multistabilities?}.  Interestingly, we show that the additional complex conjugate feedback induces distinct symmetry-breaking states as well as an explosive transition to an amplitude death state.     
	\begin{figure*}[ht!]
		\begin{center}
			\includegraphics[width=1.0\textwidth]{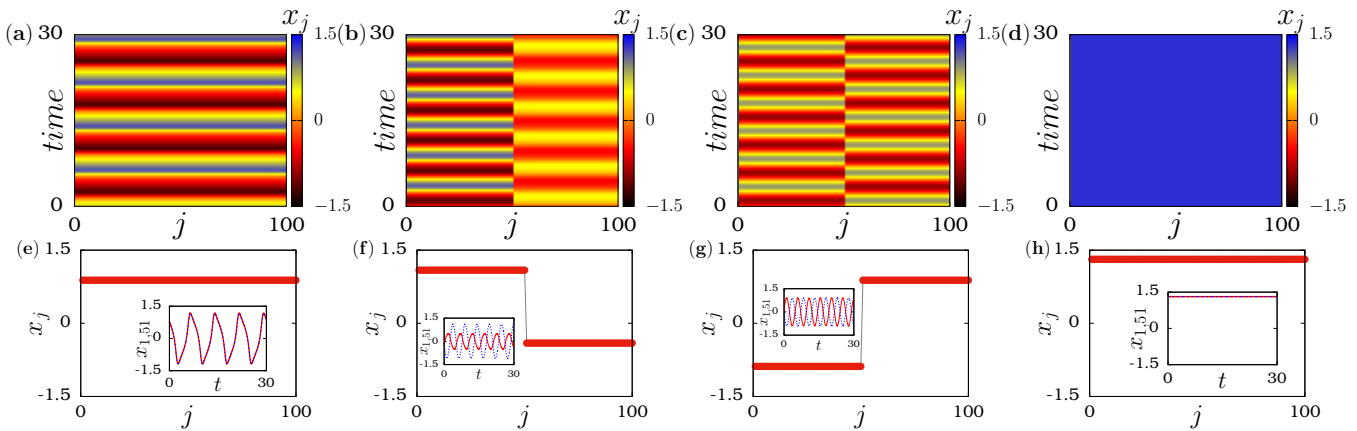}
		\end{center}
		\caption{Spatio-temporal evolution of globally coupled Stuart-Landau oscillators with additional conjugate feedback strength (a) $\eta=1.0$ (complete synchronization CS), (b) $\eta=1.3$ (symmetry breaking clusters SBC) (c) $\eta=1.5$ (out-of-phase clusters OPC), and (d) $\eta=2.0$ (nontrivial amplitude death NAD). (e)-(h) are the corresponding snapshots of $x_j$ variables and the insets represent the time-evolution of representative oscillators $x_1$ and $x_{51}$. Other parameters: $\varepsilon=0.2$,  $\Omega=1.3$, $\alpha=1.0$ and $N=100$.  }
		\label{fig:1}
	\end{figure*}
	\par In order to find out the effect of additional complex conjugate feedback factor, we considered a network of globally coupled Stuart-Landau (SL) oscillators. 
	With increasing feedback strength of the additional conjugate coupling, we find a  transition from a complete synchronization $(CS)$ to nontrivial amplitude death $(NAD)$ via symmetry breaking clusters ($SBC$) and out-of-phase clusters ($OPC$). Remarkably, an explosive transition to amplitude death is observed with the variation of the complex conjugate feedback factor. Interestingly, the hysteresis width reduces with the coupling strength of the global coupling is analyzed using the amplitude order parameter and find that the existence of first order transition with hysteresis.  
	In addition, the global dynamical transitions are illustrated in parametric space to map the occurrence of explosive transitions and multistabilities. 
	Increasing the strength or frequency of feedback enhances the multistability regions. The observed multistability regions are confirmed further by performing a bifurcation analysis and basin of attraction in the reduced model. Furthermore, the stability condition for $NAD$ is obtained using linear stability (LS) analysis.  
	
	\par   The rest of the article is organized as follows:
	The model of globally coupled SL oscillators is introduced with additional conjugate feedback in Sec. II.   Following this, the corresponding dynamical states and their transitions are analyzed in Sec.  III.  particularly, in Sec.  III and IV, we show the existence  of explosive transitions and the corresponding dynamical behaviors in the parametric spaces. Further, to confirm the existence of distinct multistability among the dynamical states, we illustrated the one-parameter bifurcation diagram and basin of attraction using the reduced model in Sec. V.   Finally, we summarize our findings in conclusion Sec. VI.    
	
	\section{The model}
	To delineate the effect of additional complex conjugate feedback, we consider a general, paradigmatic model of identical Stuart-Landau (SL) oscillators which are coupled through mean-field diffusive coupling \cite{sl1,sl2,prem_SL,sl3,sl4}.  Additionally, a complex conjugate mean-field feedback is introduced into the system of globally coupled SL oscillators; the dynamical model reads
	\begin{eqnarray}
		\label{model1} 
		\dot{w}_k=(\alpha+i \Omega -|w_k|^2)w_k +\varepsilon [\overline{w}-w_k] +\eta[{\overline{w}^*}], \\
		k=1, 2, ..., M, \nonumber
	\end{eqnarray}
	where $M$ is the chosen number of oscillators ($M=100$ in our studies). $w_k=x_k+i y_k$.   $\alpha$ is the Hopf bifurcation parameter and $\Omega$ is the system frequency.  Here $\overline{w}= \frac{1}{M}\sum_{k=1}^{M}  w_k$  is mean-field and ${\overline{w}^*}=  \frac{1}{M} \sum_{k=1}^{M} {{w}_k^*}$ or $ \frac{1}{M} \sum_{k=1}^{M} {(x_k-iy_k)}$ is the conjugate mean-field. $\varepsilon$ ($>0$) is the coupling strength of the conventional mean-field diffusive coupling and $\eta$ ($>0$) is the strength of the complex conjugate mean-field feedback. The numerical simulations are carried out using $RK_4$ algorithm with a fixed step size $h=0.01$. In the following, we will mainly explore the effect of the additional feedback factor $\eta$  on the dynamics of the system.\\ 
	
	\begin{figure*}[htb!]
		\begin{center}
			\includegraphics[width=1.0\textwidth]{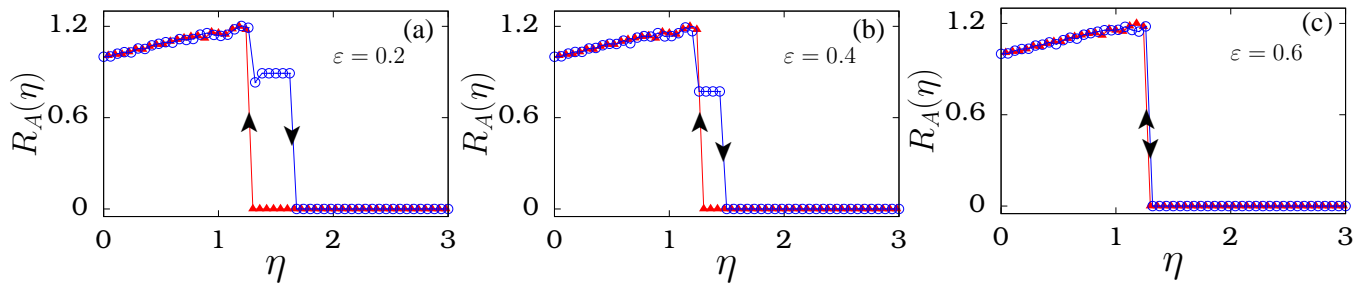}
		\end{center}
		\caption{Amplitude order parameter as a function of $\eta$ for mean-field coupling strength (a) $\varepsilon=0.2$, (b) $\varepsilon=0.4$, and $\varepsilon=0.6$. The line connecting the unfilled circles and the filled triangles denote  the forward (down arrow) and backward (up arrow) transitions, respectively.    }
		\label{fig:2}
	\end{figure*}
	\section{Dynamical states and their transitions}	
	\par {  In the absence of additional conjugate feedback, i.e., $\eta=0.0$, the coupled system Eq.~\eqref{model1} exhibits complete synchronization for all the non-zero values of coupling strength $\varepsilon$.} When the additional feedback is introduced into the coupled system, it results in the existence of two different cluster states and nontrivial amplitude death while increasing the feedback strength $\eta$. Spatiotemporal evolution and snapshots of the observed dynamical states are demonstrated in Fig.~\ref{fig:1} for $\varepsilon=0.2$ ($\Omega=1.3$). The inset in each of the snapshots represents the time evolution of representative oscillators $x_1$ and $x_{51}$. Figure~\ref{fig:1}(a) shows the emergence of complete synchronization ($CS$) for $\eta=0.2$. Due to the coherent oscillatory behavior of the $CS$ state, all the $x_j$ variables take the same value in the snapshot as depicted in Fig.~\ref{fig:1}(e). The time series trajectory of the representative oscillators $x_1$ and $x_{51}$ are also synchronized and oscillates with the same amplitudes and zero phase difference, as can be seen in the inset in Fig.~\ref{fig:1}(e). If the feedback strength is increased, we find that the oscillators in the network split into two subgroups and result in two cluster states: this is shown in Fig.~\ref{fig:1}(b) for $\eta=1.3$.  Interestingly, we noticed that the observed $x_j$ variables of the two clusters are having asymmetric values as depicted in Fig.~\ref{fig:1}(f). Such cluster states is referred as symmetry breaking cluster ($SBC$) \cite{sat_sbc1,sat_sbc2}.  From the inset in Fig.~\ref{fig:1}(f), it is clear that the oscillators $x_1$ and $x_{51}$ from the two different clusters oscillate with different amplitudes and phases. Further increase in the feedback strength results in symmetric clusters with $\pi$ phase difference. Therefore, the values take exactly equal and opposite signs (see Figs.~\ref{fig:1}(c) and \ref{fig:1}(g) for $\eta=1.5$) which is further referred as out-of-phase clusters $(OPC)$. The representative oscillators $x_1$ and $x_{51}$ from the $OPC$ state have identical amplitudes but with a $\pi$ phase difference (see the inset in Fig.~\ref{fig:1}(g)).  Further increase in $\eta$ gives rise to an oscillation quenching state. Here, we observe a nontrivial amplitude death ($NAD$) state as seen in Figs.~\ref{fig:1}(d) and \ref{fig:1}(h)). In $NAD$,  all the oscillators attain the same steady state with nontrivial coupling dependent nonzero value as is evident from  Fig.~\ref{fig:1}(h) and the representative oscillators in the inset.   In the next section, we point out the occurrence of explosive transition as a function of the feedback strength in the following.    
	
	\section{Explosive transitions}
	To illustrate the occurrence of explosive transition due to additional conjugate feedback, we have plotted the normalized average amplitude order parameter as a function of $\eta$ as shown in Fig.~\ref{fig:2}.  The order parameter in terms of  normalized average amplitude ($R_A(\eta)$)  is defined by 
	\begin{eqnarray}
		R_A(\eta)= \frac{A(\eta)}{A(0)},
		\label{model2}
	\end{eqnarray}
	where $A(\eta)$ is estimated by finding the difference between the global maxima and minima of all the oscillators averaged over a long time interval at a particular feedback strength which can be defined as $A(\eta) = (\sum_{k=1}^{M}\langle{x_{k,max}}\rangle_t-\sum_{k=1}^{M}\langle{x_{k,min}\rangle}_t)/M$. {$A(0)$ denotes the average amplitude of the oscillators in the absence of feedback.} In the death state, $R_A(\eta)$ takes null value otherwise $R_A(\eta)>0$ for  the oscillatory state. Primarily, we have plotted the amplitude order parameter for variation in $\eta$ for a fixed coupling strength  $\varepsilon=0.2$. We can note that on increasing the feedback strength the system exhibits a first-order transition, that is a transition from higher amplitudes to null value of amplitude at a critical feedback strength $\eta=1.6$ during the forward transition as shown in Fig.~\ref{fig:2}(a). Interestingly, one notes that the amplitude order parameter takes an intermediate value for the feedback strength between $1.3 \le \eta  \le 1.6$. During the reverse transition (i.e., decreasing $\eta$) results in a first-order transition from  null value of amplitude to high values at  $\eta=1.3$. Importantly, in the intermediate region, the $SBC$ or $OPC$ coexist with the $NAD$ state. Due to this coexistence,  the forward and reverse transition exhibits a hysteresis behavior.  Such first-order transition to death with hysteresis confirms the appearance of explosive death.  Furthermore, the amplitude order parameter is plotted for increased values of  the coupling strength to $\varepsilon=0.4$ and  $\varepsilon=0.6$ in Figs.~\ref{fig:2}(b) and \ref{fig:2}(c).  We also observe that increasing coupling strength $\varepsilon$ reduces the hysteresis area and suppresses it completely at a larger coupling strength. 
	\section{Dynamical transitions in the parametric spaces}
	\par
\begin{figure}
		\hspace{-0.2cm}
		\includegraphics[width=0.48\textwidth]{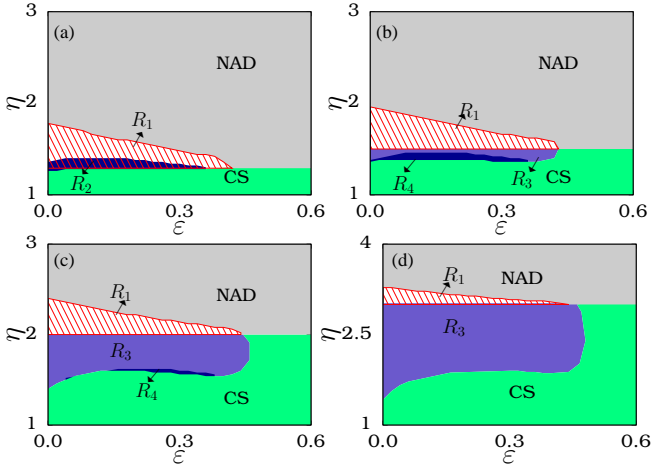}
	\caption{Two parameter diagrams in $(\varepsilon,~\eta)$ space (a) for $\Omega=1.3$,  (b) for $\Omega=1.5$, (c)  for $\Omega=2.0$, and (d) for $\Omega=3.0$.  $CS$ and $NAD$ are the complete synchronization and nontrivial amplitude death states, respectively.  $R_1,~R_2,~R_3$ and $R_4$ are the multistability regions of  $OPC-NAD$, $SBC-NAD$, $OPC-CS$,  and $CS-SBC$, respectively.  }
	\label{fig:3}
\end{figure}

 Furthermore, for a more clear understanding of the explosive transitions and multistabilities, we analyze the dynamical behavior of the system specified by Eq.~\ref{model1} through parametric spaces in detail.  Initially, to understand the dynamical transitions in the parametric space, the two-parameter diagrams are plotted in $(\varepsilon,\eta)$ parametric space in Fig.~\ref{fig:3} by fixing the frequency at four different values, namely $\Omega=1.3,~1.5,~2.0$ and $\Omega=3.0$.  The {striped patterns} indicate the hysteresis area. In Fig.~\ref{fig:3}(a) for $\Omega=1.3$, at lower feedback strength with increasing  coupling strength, the system (1) exhibits a complete synchronization behavior for the full range of coupling strength. Increasing the feedback strength, the system (1) gives rise to multistabilities between $SBC$  and $NAD$ as well as $OPC$  and $NAD$ states represented by the regions $R_1$ and $R_2$, respectively. Increasing the feedback to larger values, one finds a suppression of the multistability regions and one finds $NAD$ regions in the entire parametric space. On increasing the frequency to $\Omega=1.5$, we observe that the $CS$ region is increased.  Subsequently, $R_2$ region is suppressed with the onset of additional multistability regions $R_3$ and $R_4$.  In these regions, $CS$ coexists with $OPC$ and $SBC$ states, respectively. Increasing the frequency to $\Omega=1.5$  and $\Omega=2.0$ further, one finds the $R_3$ region gets widened, while the region $R_4$ is  reduced. 
	
\begin{figure}
	\hspace{-0.5cm}
	\includegraphics[width=0.5\textwidth]{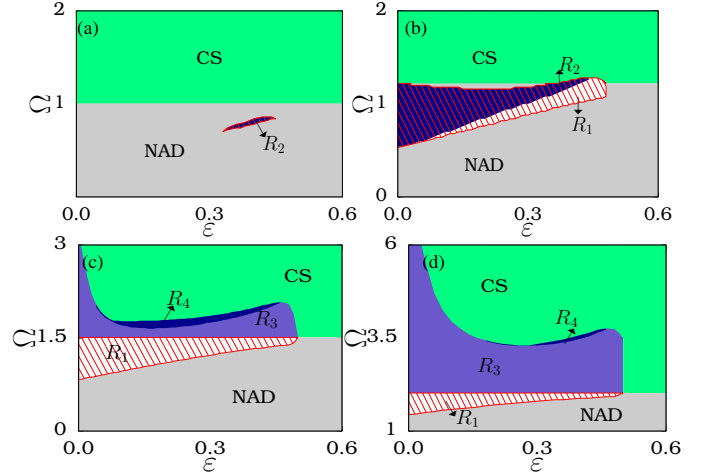}
	\caption{Two parameter diagram in $\varepsilon,~\Omega$ space for (a) $\eta=1.0$,  (b)  $\eta=1.2$, (c)  $\eta=1.5$, and (d)  $\eta=2.0$.  $CS$ and $NAD$ are the complete synchronization and nontrivial amplitude death, respectively.  $R_1,~R_2,~R_3$ and $R_4$ are the multi-stability regions of  $OPC-NAD$, $SBC-NAD$, $OPC-CS$,  and $CS-SBC$, respectively. }
	\label{fig:4}
\end{figure}

\par In order to illustrate the occurrence of multistabilities and explosive transition due to the additional conjugate feedback, the two-parameter diagrams are also plotted in Fig.~\ref{fig:4} by fixing the feedback strength at different values. For feedback strength $\eta=1.0$, an increasing frequency  manifests in a transition from $NAD$ to $CS$ for all values of coupling strengths (see Fig.~\ref{fig:4}(a)). Interestingly, we noticed that the  onset of  $SBC$ region at certain  critical values of $\varepsilon$ and $\Omega$ in the $NAD$ region (denoted as $R_2$).  Increasing the feedback strength to  $\eta=1.2$ (depicted in Fig.~\ref{fig:4}(b)),  one finds that the system (1) gives rise to wider region of $R_2$.  Also, we noticed the existence of $R_1$ region at lower frequencies.  On increasing the feedback strength to $\eta=1.5$, one finds that the $R_2$ region diminishes.  Eventually, we observe additional two different multistability regions $R_3$ and $R_4$ (see  Fig.~\ref{fig:4}(c)). Upon increasing  the feedback as shown in  Fig.~\ref{fig:4}(d), one observes an enlarged $R_3$ region.  From the above  observations, it is clear that the additional conjugate feedback results in interesting multistability among the dynamical states in the parametric space.   
\section{Dynamical behaviors through reduced model}

\begin{figure}
	\hspace{-0.91cm}
	\includegraphics[width=0.52\textwidth]{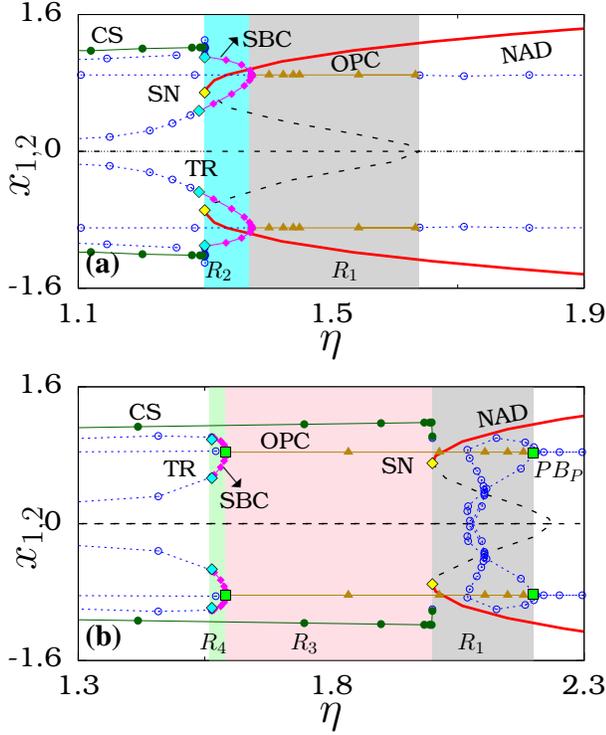}
\caption{ One parameter bifurcation diagram  for (a) $\varepsilon=0.2$,  $\Omega=1.3$ and (b)  $\varepsilon=0.3$,  $\Omega=2.0$.  $R_1,~R_2,~R_3$ and $R_4$ are the multi-stability regions of $OPC-NAD$, $SBC-NAD$, $OPC-CS$, and $CS-SBC$, respectively.  The line connecting by filled circles, diamonds, and triangles denotes the stable $CS$, $SBC$, and $OPC$ state, respectively.  $CS$, $SBC$, and $OPC$ are the complete synchronization, symmetry-breaking clusters, and out-of-phase clusters, respectively.  $NAD$ represents the nontrivial amplitude death state. {\bf $TR,~SN$ and $PB_s$ denote the torus, saddle-node, and pitchfork fork bifurcation points, respectively. } The solid red line represents the stable steady state. The open circles and dashed lines indicate the unstable steady state and unstable oscillations, respectively. }
\label{fig:5}
\end{figure}

{ In order to understand the genesis of the explosive transition and multistability, we perform a bifurcation analysis and basin stability analysis in the following. } 
For this purpose, we reduce our considered system (Eq.~\ref{model1}),  by splitting the network into two groups $w_k=W_1$ for $k=1, 2, ...M(1-p)$ and  $w_k=W_2$ for $k=M(1-p)+1, ..., M$, where $p=0.5$.  The corresponding dynamical equations can be written as
\begin{eqnarray}
	\dot{W_1}&=&(\alpha+i\Omega-|W_1|^2)W_1 +\frac{\varepsilon}{2} (W_2-W_1)+\frac{\eta}{2}{{W}^*},\nonumber\\  
	\dot{W_2}&=&(\alpha+i\Omega-|W_2|^2)W_1 +\frac{\varepsilon}{2}(W_1-W_2)+\frac{\eta}{2}{{W}^*},
	\label{model3}
\end{eqnarray}
where ${W^*}={W_1^*}+{W_2^*}$.
To show the dynamical transitions and multistability, primarily we have plotted the one-parameter bifurcation diagram (using XPPAUT \cite{xpp}) by fixing $\varepsilon=0.2$,  $\Omega=1.3$ in Fig.~\ref{fig:5}(a).   Increasing the feedback strength shows a  direct transition from a homogeneous oscillatory state to a homogeneous steady state that is $CS$ to $NAD$ through saddle-node bifurcation $(SN)$ at the feedback strength $\eta=1.3$.   In addition for a particular value of feedback ($\eta=1.29$) the onset of the $SBC$ takes place through Torus $(TR)$ bifurcation.  The amplitudes of the $SBC$ cluster have different values which decrease while the value of $\eta$ increases and attains the same values of amplitude where it transits to $OPC$. Further, the $OPC$ state loses its stability at a critical feedback strength $\eta=1.7$ and attains the $NAD$ state.   The shaded regions represent the multistability regions $R_1$ and $R_2$, respectively.   

\par In addition, to delineate other multistability zones, we set the parameters at $\varepsilon=0.3$ and $\Omega=2.0$ and depict a one parameter diagram, as shown in Fig.\ref{fig:5}(b). The bifurcation diagram clearly shows that on increasing the value of $\eta$ results in the birth of $SBC$ through the $TR$ bifurcation, which coexists with $CS$ shown by $R_4$. As $\eta$ increases, the $SBC$ state transits to the $OPC$ state. The $CS$ state coexists with the $OPC$ state in the area of multistability region $R_3$. If the feedback is increased further, the coexistence of the $OPC$ and $NAD$ regions (denoted by $R_1$) is revealed.     

{ Further, one can also determine the stability criterion for the oscillation quenching state using the reduced model.}
In order to find the stability of the observed nontrivial amplitude death state, we first estimate the fixed point  $(x_{1},y_{1},x_{2},y_{2})=(\pm x^{*},\pm y^{*},\pm x^{*},\pm y^{*})$ using Eq.~(\ref{model3}),
\begin{eqnarray}
\label{fp} 
x^{*}&=& \mp \frac{\sqrt{\eta+\alpha-\frac{\Omega^2}{\eta}-\frac{\tilde{\eta}}{\eta^2}}}{\sqrt{2}}   \nonumber\\
y^{}&=& \mp \frac{\eta^3+\eta^2\alpha-\tilde{\eta}}{\eta(\eta+\alpha)\Omega} x^*
\end{eqnarray}
where $\tilde{\eta}=\sqrt{\eta^2(\eta+\alpha)^2(\eta^2-\Omega^2)}$.  Further, by finding the Jacobian matrix of Eq.~(\ref{model3}) the eigenvalues  are obtained which are expressed as 
\begin{eqnarray}
\label{eigen}
\lambda_{1,2}&=& \alpha-2r^* \mp \sqrt{r^{*2}-\Omega^2}, \nonumber \\
\lambda_{3,4}&=& \alpha-2r^* \mp \sqrt{r^{*2}-2d^*+\eta^2-\Omega^2},   
\end{eqnarray}
where $r^*=x_1^{*2}+y_1^{*2}$ and  $d^*=y_1^{*2}-x_1^{*2}$. The stability condition for $NAD$ is identified by equating the real part of the eigenvalues to zero, and is obtained as $\eta_{_{SN}}=\Omega$.  

\begin{figure*}[htb!]
\begin{center}
	\includegraphics[width=0.8\textwidth]{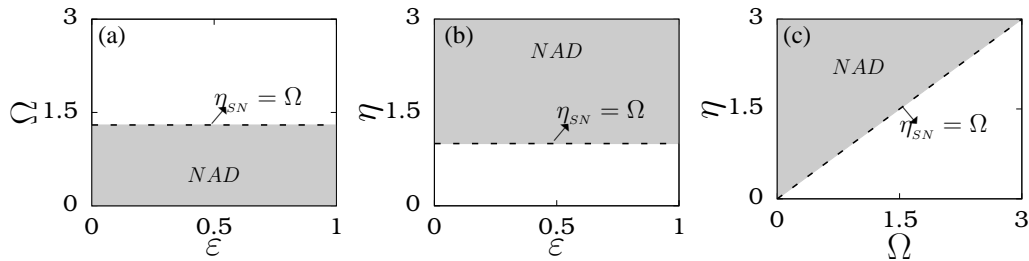}
\end{center}
\caption{Stable $NAD$ regions in (a) $(\varepsilon,\Omega)$ space for $\eta=1.0$,  (b)  $(\varepsilon,\eta)$ space for $\Omega=1.0$,  and  (c) $(\Omega,\eta)$  space  for $\varepsilon=1.0$.  The dashed line in each figure indicates the stability curve which separates $CS$ and $NAD$ states.    }
\label{fig:7}
\end{figure*}

\par  Using the stability condition $\eta_{SN}=\Omega$, we present the stable $NAD$ region in different parametric spaces in Fig.\ref{fig:7}. The stability curve for the $NAD$ region, which separates the $CS$ and $NAD$ regions, is shown by the dashed line in each diagram.  From Fig.\ref{fig:7}(a), it is clear that  for $\eta=1.0$, the stable boundary exists at $\Omega=1$ and  the stability region occurs for $\Omega\le1.0$ for all values of $\varepsilon$. Analogously, from   Fig.\ref{fig:7}(b) in $(\varepsilon,\eta)$ space, the stability region occurs at  $\eta=1$ and $\eta\ge1.0$ for all values of $\varepsilon$ and $\Omega=1$. Furthermore, for any value of $\varepsilon$,  stable $NAD$ region  in  $(\Omega,\eta)$  space exists when   $\eta_{_{SN}}\ge\Omega$ as seen in Fig.~\ref{fig:7}(c).  It is also noticed that the fixed points and stability conditions of $NAD$ are independent of the coupling strength ($\varepsilon$).  Hence, the boundary does not change with respect to   $\varepsilon$.  It can be seen that the stable boundaries of $NAD$ region matches well with the numerically obtained $NAD$ region (cf. Fig.~\ref{fig:4}(a) and Fig.~\ref{fig:3}(a) ).

\par
\begin{figure*}[htb!]
\begin{center}
	\includegraphics[width=1.0\textwidth]{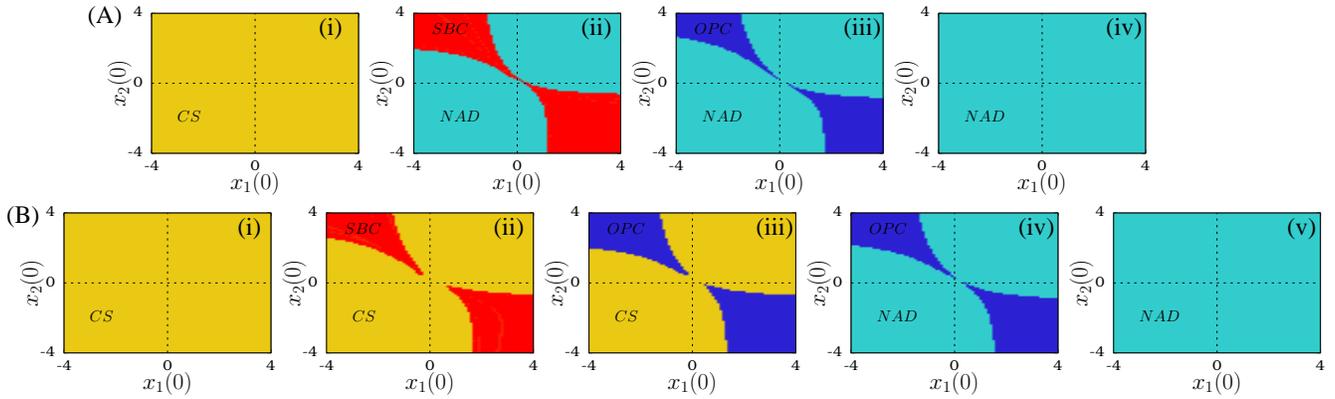}
\end{center}
\caption{Basin of attraction by fixing the initial states at $y_1(0)=0.5$ and $y_2(0)=0.45$ and varying  $x_1(0)$ and   $x_2(0)$ for panel  A: (i) $\eta=1.1$,  (ii) $\eta=1.37$,  (iii) $\eta=1.5$  and (iv) $\eta=1.8$.  For panel B:  (i) $\eta=1.3$,  (ii) $\eta=1.58$,  (iii) $\eta=1.8$, (iv) $\eta=2.0$, and  (v) $\eta=2.3$.  The other parameters for the panels (A) and (B) are fixed as corresponding to  Figs.~\ref{fig:5}(a) and \ref{fig:5}(b), respectively.}
\label{fig:6}
\end{figure*}

 For validating the existence of bistability among the dynamical states, the basin of attraction is plotted  in Fig.~\ref{fig:6} using Eq.~(\ref{model2}) by fixing the initial state of $y_1(0)$ and $y_2(0)$ and by varying $x_1(0)$ and $x_2(0)$ \cite{basin}. Firstly, the upper panel  A is plotted by fixing  $\varepsilon=0.2$,  $\Omega=1.3$, and for different values of feedback strength.   For $\eta=1.1$, We can observe that the entire initial state space is filled with a complete synchronization state as shown in Fig.~\ref{fig:6}A(i). If the feedback strength is increased to  $\eta=1.37$, we notice that some of the asymmetric initial states favor $SBC$ and the remaining initial states favor the $NAD$ state (see Fig.~\ref{fig:6}A(ii)).  Here, the asymmetric initial states represented by  $x_1(0)$ reach positive initial states while $x_2(0)$ attains negative values or vice versa. Thus, Fig.~\ref{fig:6}A(ii) provides evidence for the co-existence of $SBC$ and $NAD$. Further, on increasing the feedback strength to $\eta=1.5$, one observes that $SBC$ gets wiped out by $OPC$  (see Fig.~\ref{fig:6}A(iii)).   Upon increasing the feedback strength further  we find that the entire initial state space is occupied by $NAD$ state  (see Fig.~\ref{fig:6}A(iv)). Similarly, panel B is plotted for $\varepsilon=0.3$, $\Omega=2.0$, which corresponds to Fig.~\ref{fig:5}(b). For $\eta=1.3$, we observed that the entire basin is filled with $CS$ state, as shown in Fig.~\ref{fig:6}B(i). When the feedback strength is increased to $\eta=1.58$, certain asymmetric initial states result in the $SBC$ state, while the others lead to the $CS$ state (see Fig.~\ref{fig:6}B(ii)). If the feedback is increased to $\eta=1.8$, the $SBC$ transits to the $OPC$ state, which coexists with the $CS$ state, as shown in Fig.~\ref{fig:6}B(iii). When the feedback strength is increased further ($\eta=2.0$), the $CS$ in the basin is completely suppressed. Finally, the $NAD$ state occupies the basin  instead of $CS$ state (see Fig.~\ref{fig:6}B(iv)). As the feedback gets stronger, it acquires $NAD$ for the entire basin, as shown in Fig.~\ref{fig:6}B(v). From these observations, it is clear that the onset of multistability among the dynamical states is the major impact of feedback strength. 

\section{Conclusions} 
\par The feedback factor is known to be vital for reviving oscillation death or restoring rhythmicity in a dynamical system. The purpose of this study is to analyze whether the feedback component may lead to explosive death and multistability. To accomplish this, we have considered a system of globally coupled Stuart-Landau oscillators with conjugate feedback. We first looked at dynamical transitions by increasing the feedback factor. We discovered that the transition from complete synchronization ($CS$) to nontrivial amplitude death ($NAD)$ occurs via symmetry breaking clusters $(SBC)$ and out-of-phase clusters $(OPC)$. The emergence of explosive transitions is demonstrated using the amplitude order parameter as a function of feedback strength. The occurrence of first-order transitions is linked to hysteresis behavior. 
The region of multistable states is found to be governed by the interplay of feedback factors and the natural frequency of the coupled oscillators. 
For a better understanding of the observed scenarios, we have considered a reduced model and carried out a detailed bifurcation analysis and a linear stability analysis, which are in well accordance with the numerical results of the extended system. The obtained results imply that the globally coupled oscillators exhibit an explosive death and distinct multistable states as a result of additional conjugate feedback. {  Additionally, the proposed work raises many open problems. For instance, extending our study to different topological structures, such as scale-free, small world, and others, is a practical realization. Feedback is a common way to restore dynamism to the degraded dynamical units in a complex network, and it is also crucial to investigate the effects of different feedbacks.  }

Our findings may offer insight on the impact of additional feedback in many biological systems where feedback appears naturally to regulate physiological mechanisms \cite{bio1,bio2} and in engineering systems where often feedback is wishfully introduced for the control purpose (e.g., in phase-locked loops and laser systems \cite{pll,laser-sch}).

\section*{Acknowledgments}
DP gratefully acknowledge this work is funded by the Center for Nonlinear Systems, Chennai Institute of Technology (CIT), India, vide funding number CIT/CNS/2022/RP-016. ML is supported by a DST-SERB National Science Chair position (Ref. no. NSC/2020/000029).

\end{document}